\documentclass[12pt,twoside,a4paper]{article}
\usepackage[dvips]{epsfig}
\begin{document}

\begin{titlepage}
\vspace*{-1.0cm}
{\flushright LC-PHSM-2000-052 \\}

\vspace{1.50cm}

\begin{center}
{\huge \bf Study of $\gamma \gamma$ Background \\ 

\vspace{0.15cm}

in $e^+e^- \rightarrow W^+W^- \nu \bar \nu \rightarrow H^0 \nu \bar \nu$
Events \\

\vspace{0.15cm}

at the {\sc Tesla} $e^+e^-$ Linear Collider}

\vspace{1.50cm}

{\large M. Battaglia}\\
Dept. of Physics, University of Helsinki (Finland)\\
and CERN, Geneva (Switzerland)

\vspace{0.50cm}

{\large D. Schulte}\\
CERN, Geneva (Switzerland)
\end{center}

\vspace{2.50cm}

\begin{abstract}
The effect of the overlap of $\gamma \gamma \rightarrow {\mathrm{hadrons}}$
to $H \nu \bar \nu$ events has been studied for the case of the {\sc Tesla}
$e^+e^-$ linear collider at $\sqrt{s}$ = 350~GeV. It was found that,
due to the significant bunch length and the track extrapolation accuracy 
provided by the Vertex Tracker, the $\gamma \gamma$ background to physics 
events can be substantially reduced, with moderate loss in reconstruction 
efficiency, by a combination of kinematical and vertex topology observables. 
The remaining background, being confined to very forward hadron production, 
does not significantly interfere with the event reconstruction.

\end{abstract}

\end{titlepage}

\section{Introduction}

Two photon interactions are characterised by a cross section that is several
orders of magnitude larger compared to those typical for boson and fermion 
production in $e^+e^-$ collisions. 
At a $e^+e^-$ linear collider,  a high rate of $\gamma \gamma$
collisions arises from photons radiated in the electro-magnetic interactions 
of the colliding beams. While interesting for themselves, products of 
$\gamma \gamma$ interactions may overlap with those from a $e^+e^-$
interaction creating a source of potential confusion for the event 
reconstruction.

Due to the large luminosity per bunch crossing $L = 2.2 \times 
10^{-3}$~nb$^{-1}$~BX$^{-1}$ achieved at {\sc Tesla}, the rate of 
$\gamma \gamma \rightarrow$ hadrons overlap to a $e^+ e^-$ collision is
0.09 - 0.13 $\times n_b$, where $n_b$ is the number of bunches integrated 
within a read-out cycle of the relevant detector systems.
The considerably high rate of this event overlap has been considered as
a possible limiting factor in the precise study of processes whose
production cross section is peaked in the forward region and  experimental 
signature includes missing energy~\cite{yamashita}. In this report, the 
identification of two photon interactions overlapped to 
$e^+e^- \rightarrow W^+W^- \nu \bar \nu \rightarrow H^0 \nu \bar \nu$ 
Higgs production via t-channel $WW$ fusion is discussed. 
This channel offers a good benchmark process. An accurate
determination of the Higgs production cross-section by the $WW$ fusion
process is important for a determination of the Higgs total width in 
conjunction with the $H^0 \rightarrow WW$ decay branching 
fraction~\cite{klaus}. Further,
for a large interval of $M_H$ and $\sqrt{s}$ values this process is the 
dominant Higgs production mechanism in $e^+e^-$ annihilations. 
The $\gamma \gamma$
overlap may distort both the reconstruction efficiency and the 
jet flavour tagging performance. Therefore it is of great importance to
ensure that this $\gamma \gamma$ background can be effectively rejected
in the Higgs studies.

\section{$\gamma \gamma$ Simulation}

Two photon events have been generated by the {\sc Guinea~Pig} 
program~\cite{c:thesis}.
Beam parameters for 350~GeV centre-of-mass were derived from the ones
for 500~GeV by keeping the beta-functions and emittances constant.
The hadronic background has been modelled following Schuler and
Sj\"ostrand~\cite{c:had0} using the most conservative proposed 
parametrisation of the cross section:
\begin{equation}
\sigma_{H,3}=211{\mathrm{nb}}\cdot \left(\frac{s}{{\mathrm{GeV}}} 
\right)^{0.0808}
+297{\mathrm{nb}}\cdot \left(\frac{s}{{\mathrm{GeV}}}\right)^{-0.4525}
\label{e:had0}
\end{equation}
The energies of two colliding photons have been stored in a file with their 
relevant probability. {\sc Pythia}~\cite{c:pythia} has then been used to 
generate the hadronic events.
These events have been stored in a database from where they could be added to 
the main event stored in the {\sc Pythia} common block.

The final state hadronic system has been appended to the event record
of $e^+e^- \rightarrow W^+W^- \nu \bar \nu \rightarrow H^0 \nu \bar \nu$
events generated using the {\sc Pythia} program.

\section{Event Analysis and $\gamma \gamma$ Tagging}

\begin{figure}[h!]
\begin{center}
\begin{tabular}{c c}
\epsfig{file=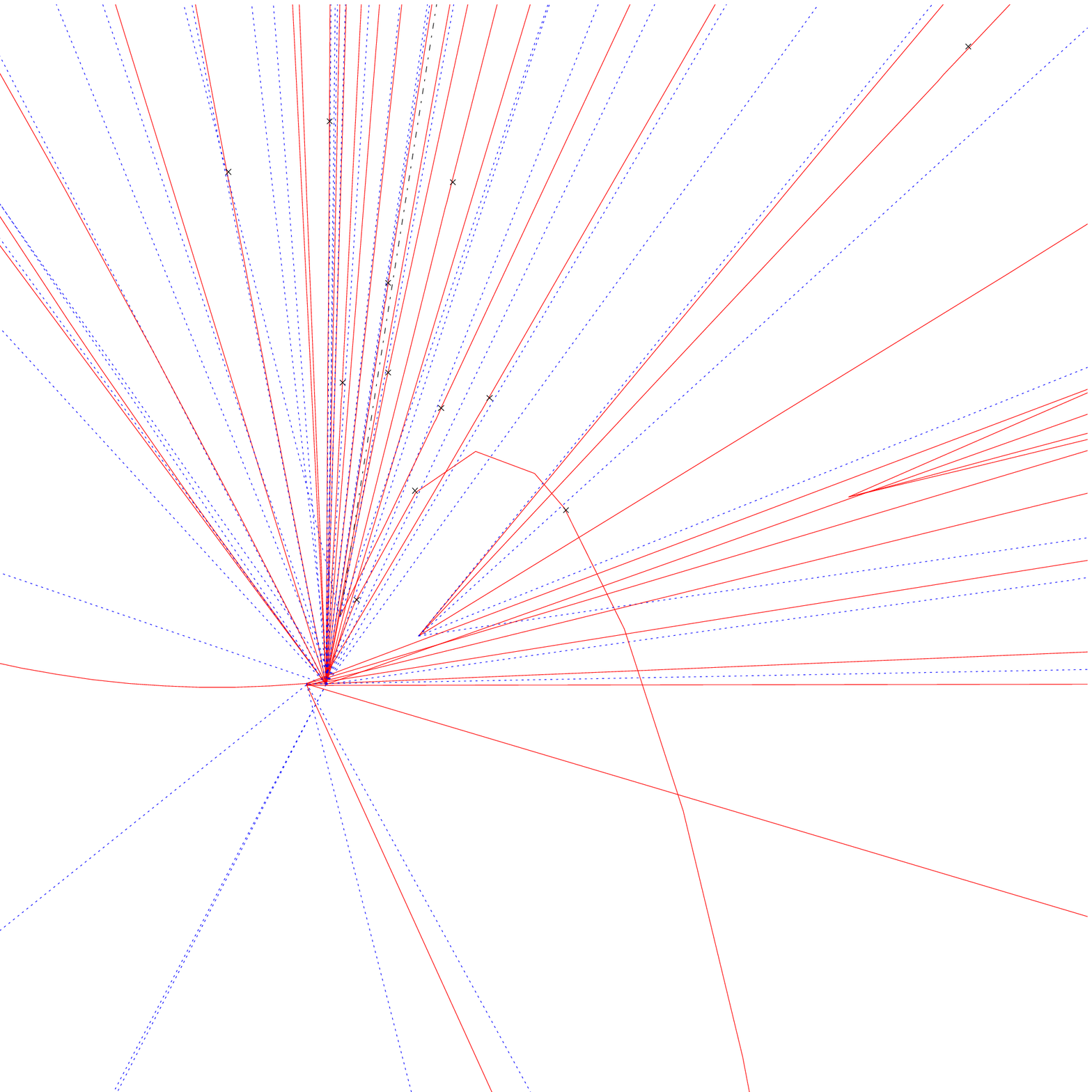,width=7.0cm} &
\epsfig{file=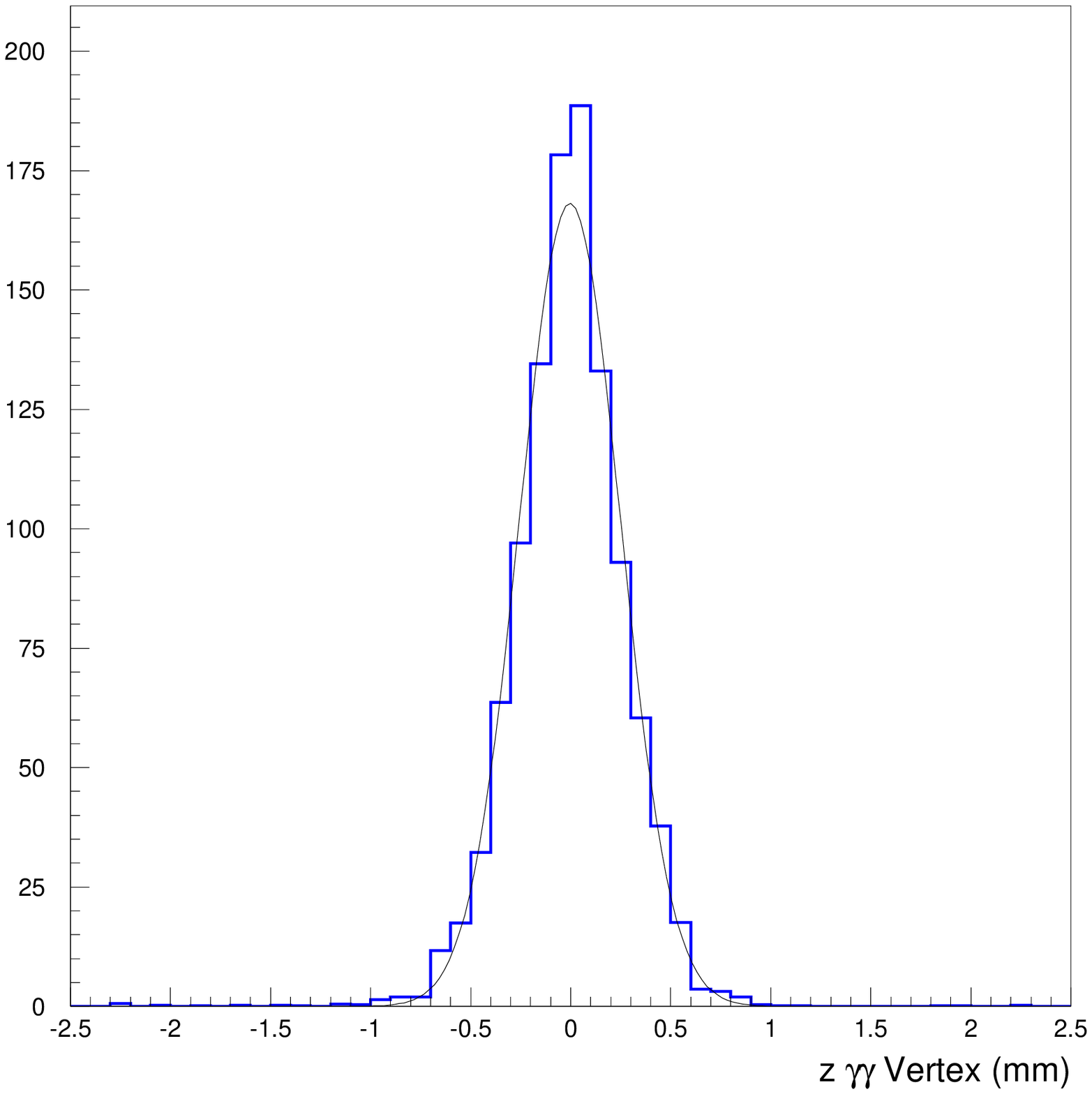,width=8.0cm} \\
\end{tabular}
\end{center}
\caption[]{\sl Left: Display of a $e^+e^- \rightarrow W^+W^- \nu \bar \nu 
\rightarrow H^0 \nu \bar \nu$, $H^0 \rightarrow b \bar b$ decay with an 
overlapped $\gamma \gamma$ event showing the secondary
vertex structure of the decays of $b$ hadrons and the longitudinal 
displacement of the $\gamma \gamma$ production point w.r.t. the $e^+e^-$
interaction.
Right: The distribution of the $\gamma \gamma$ vertex position along the 
beam axis.}
\label{fig:evt}
\end{figure}

$\gamma \gamma \rightarrow$~hadrons events are characterised by forward 
hadronic activity and by a production vertex contained in the colliding
bunch envelope but distinct from that of the $e^+e^-$ events. Since at
{\sc Tesla} the bunch size is 300~$\mu$m in length but only few nm in the
vertical and horizontal projections, $\gamma \gamma \rightarrow$ hadrons 
events are effectively displaced only in the longitudinal direction
(see Figure~\ref{fig:evt}).

$e^+e^- \rightarrow W^+W^- \nu \bar \nu \rightarrow H^0 \nu \bar \nu$
events have been generated for $M_H$ = 120~GeV/$c^2$ and $\sqrt{s}$ =
350~GeV, hadrons from $\gamma \gamma$ interactions have been added and 
the resulting events passed through the {\sc Geant} simulation of 
the {\sc Tesla} detector tracking system. It has been assumed that the 
Vertex Tracker can identify single bunches within a {\sc Tesla} pulse, as
in the case of hybrid pixel sensors~\cite{pixels}, corresponding to an
average probability for $\gamma \gamma$ event overlap of 0.10.

\begin{figure}[ht!]
\begin{center}
\begin{tabular}{c c}
\epsfig{file=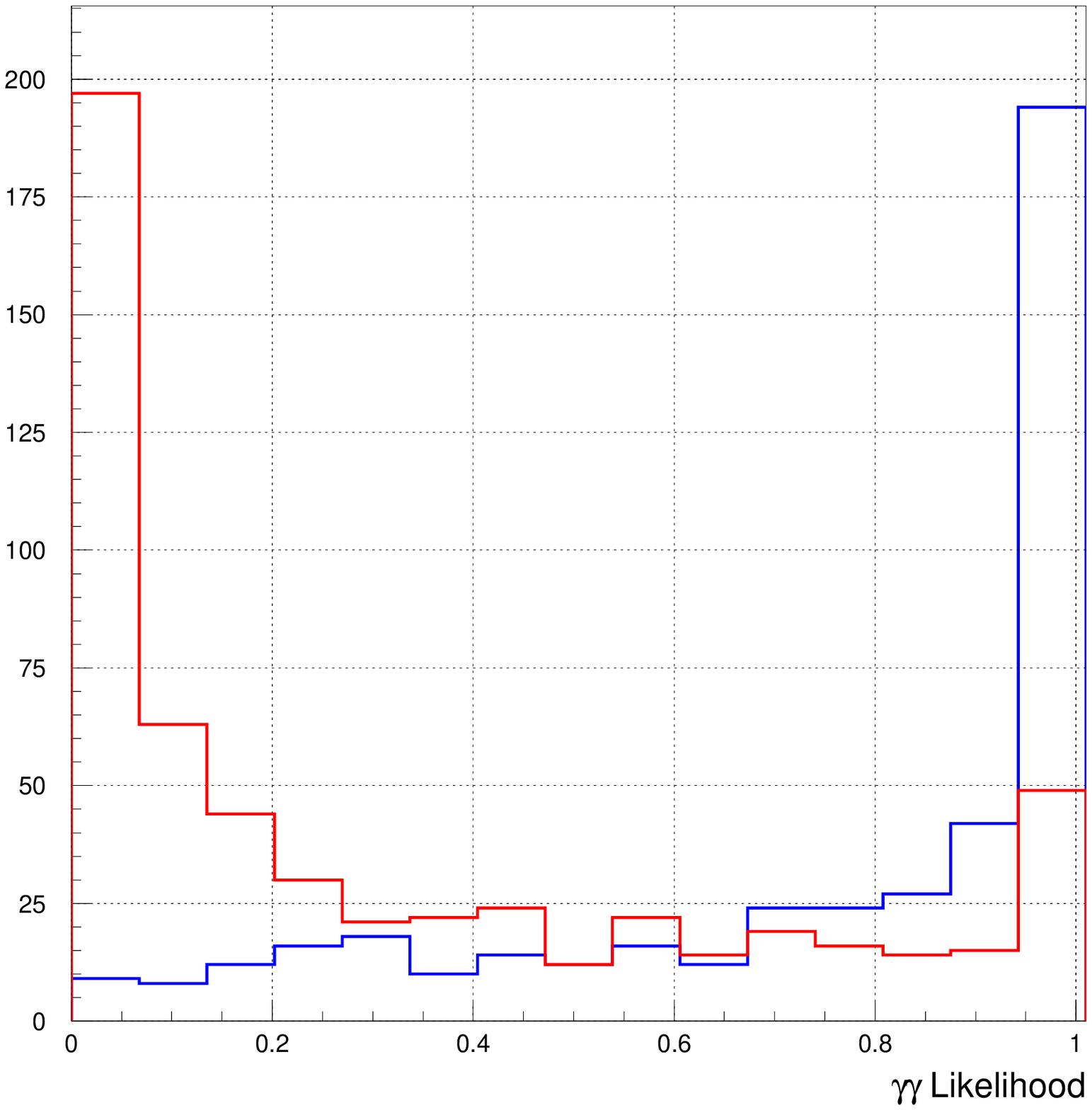,width=7.0cm,height=6.5cm} & 
\epsfig{file=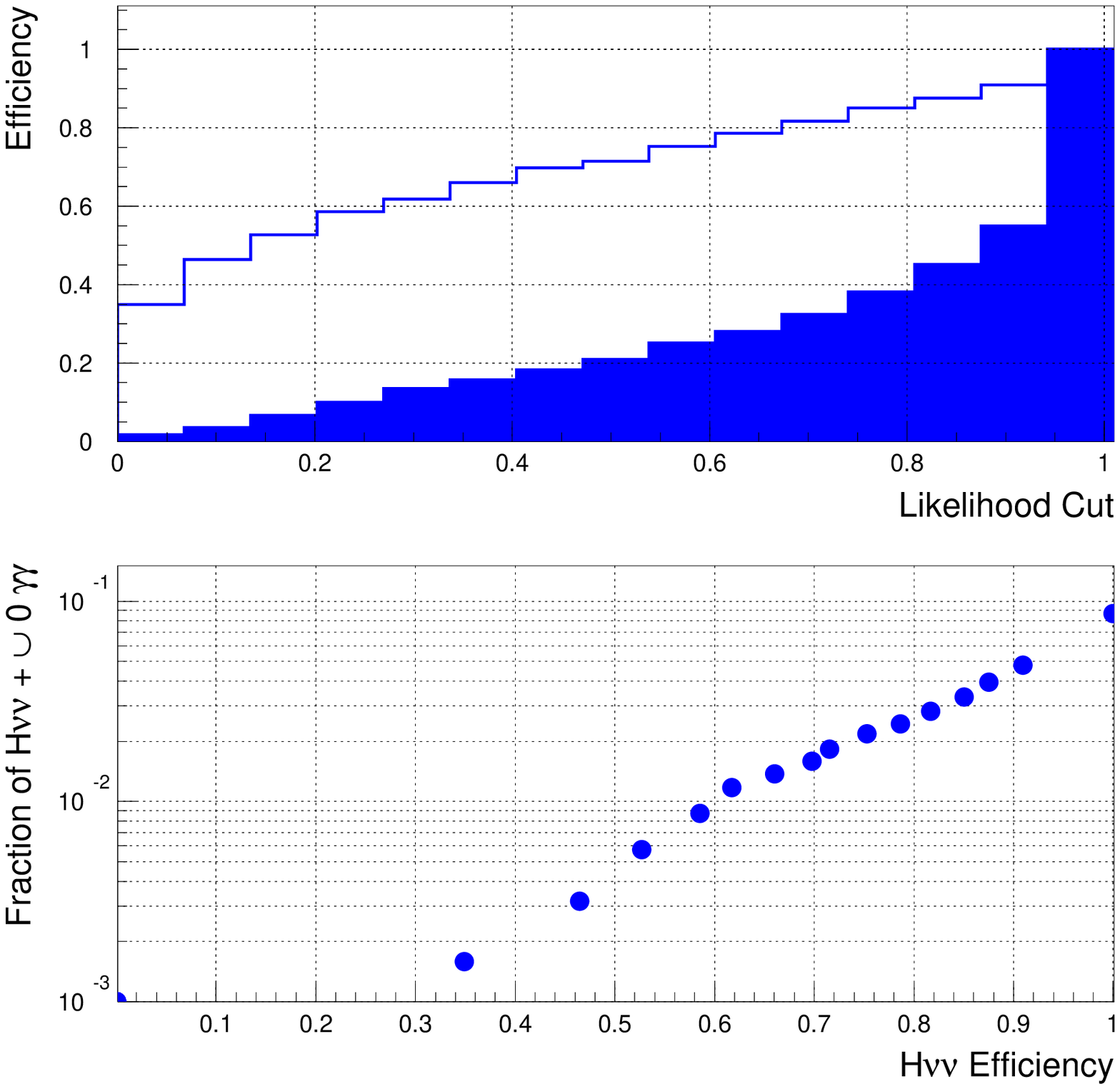,width=7.0cm,height=6.5cm} \\
\end{tabular}
\end{center}
\caption[]{\sl Left: The $\gamma \gamma$ likelihood for $H \nu \bar \nu$ and 
$H \nu \bar \nu + \ge~1 \gamma \gamma$ events. Right: The efficiency for
retaining a $H \nu \bar \nu$ event as a function of $\gamma \gamma$ likelihood
cut value (upper plot) and  the fraction of tagged $H \nu \bar \nu + 
\ge~1 \gamma \gamma$ events shown as a function of the efficiency 
(lower plot).}
\label{fig:eff}
\end{figure}

The tagging algorithm developed for identifying and rejecting overlapped
$\gamma \gamma \rightarrow$ hadrons event is based on the topological 
and kinematical characteristics of the particles detected in the forward
regions.
The forward hadronic system was reconstructed from the particles detected
at polar angles $0.80 < |\cos \theta| < 0.96$. The displacement of the 
production vertex was verified from the reconstructed impact parameter
w.r.t the event primary vertex reconstructed using the primary particles
detected in the central region defined as $|\cos \theta| < 0.80$. 
In order to distinguish the 
$\gamma \gamma$ products, displaced only in the $z$ component, from decay
products of short-lived hadrons, such as in $H^0 \rightarrow b \bar b$,
both the $R-\phi$ and $z$ impact parameter components have been studied.
A track extrapolation resolution, $\sigma^{IP}_{R-\phi, z}=
\sqrt{(5.7~\mu m)^2 + \frac{(13.0~\mu m)^2}{p_t^2}}$, better than a tenth of 
the {\sc Tesla } bunch length can be achieved by the use of a high precision 
silicon Vertex Tracker. This provides a good discrimination of primary 
particles, secondaries from short-lived hadron decays and hadrons from 
$\gamma \gamma$ interactions.
Eight variables sensitive to the kinematical and topological properties of the
reconstructed forward hadronic system have been used to define a likelihood
variable used for the tagging:
\begin{enumerate}
\item{Prob($IP_z$) for forward tracks with Prob($IP_{R-\phi}$) $> 0.05$;}
\item{Number of hemispheres with forward activity;}
\item{$z$ position of reconstructed detached Vertex (1-D);}
\item{fraction of forward tracks with Prob($IP_z$)$< 0.05$ and
Prob($IP_{R-\phi}$) $>0.05$;}
\item{Number of hadronic jets;}
\item{Ranking of most forward jet in energy ordering;}
\item{$|\cos \theta|$ of most forward jet;}
\item{Invariant mass of the forward hadronic system;}
\end{enumerate}  
A global $\gamma \gamma$ likelihood $L_{\gamma \gamma}$ was defined from the 
response of the selected variables as:
\begin{eqnarray}
L_{\gamma \gamma} = \frac{\prod_{i=1,8} F_{i}^{\gamma \gamma}(x_i)} 
{\prod_{i=1,8} F_{i}^{\gamma \gamma}(x_i) + 
(\prod_{i=1,8} (1 - F_{i}^{\gamma \gamma}(x_i)))}
\end{eqnarray}
where $F_{i}^{\gamma \gamma}(x_i)$ is the probability for an event with 
$\gamma \gamma$ background to give a value $x_i$ for the $i^{th}$ observable.
This variable peaks at one (zero) for $H \nu \bar \nu$ events with at least one
$\gamma \gamma$ overlap (pure $H \nu \bar \nu$ events) as shown in 
Figure~\ref{fig:eff}.
At 80\% $H \nu \bar \nu + 0~\gamma \gamma$ efficiency, corresponding to a cut
$L_{\gamma \gamma} \le 0.65$, the fraction 
of events with $> 0~\gamma \gamma$  accepted in the analysis is $\le 0.025$.
The residual events, left after a cut on the $\gamma \gamma$ likelihood 
variable, are concentrated in very forward region beyond the acceptance of
the Vertex Tracker.

In order to check the effect of the remaining $\gamma \gamma$ background on
the event reconstruction, the di-jet invariant mass $M_{jj}$ and their recoil 
mass $M_{recoil}$ have been computed for $H \nu \bar \nu$, $H \nu \bar \nu + 
\gamma \gamma$ events and for the events satisfying the anti-$\gamma \gamma$ 
cut $L_{\gamma \gamma} \le 0.65$, corresponding to 80\% efficiency. 
Hadronic jets
have been reconstructed using the Durham algorithm, only events with at most
two jets in the central regions have been retained. 

\newpage

\begin{figure}[hb!]
\begin{center}
\begin{tabular}{c c}
\epsfig{file=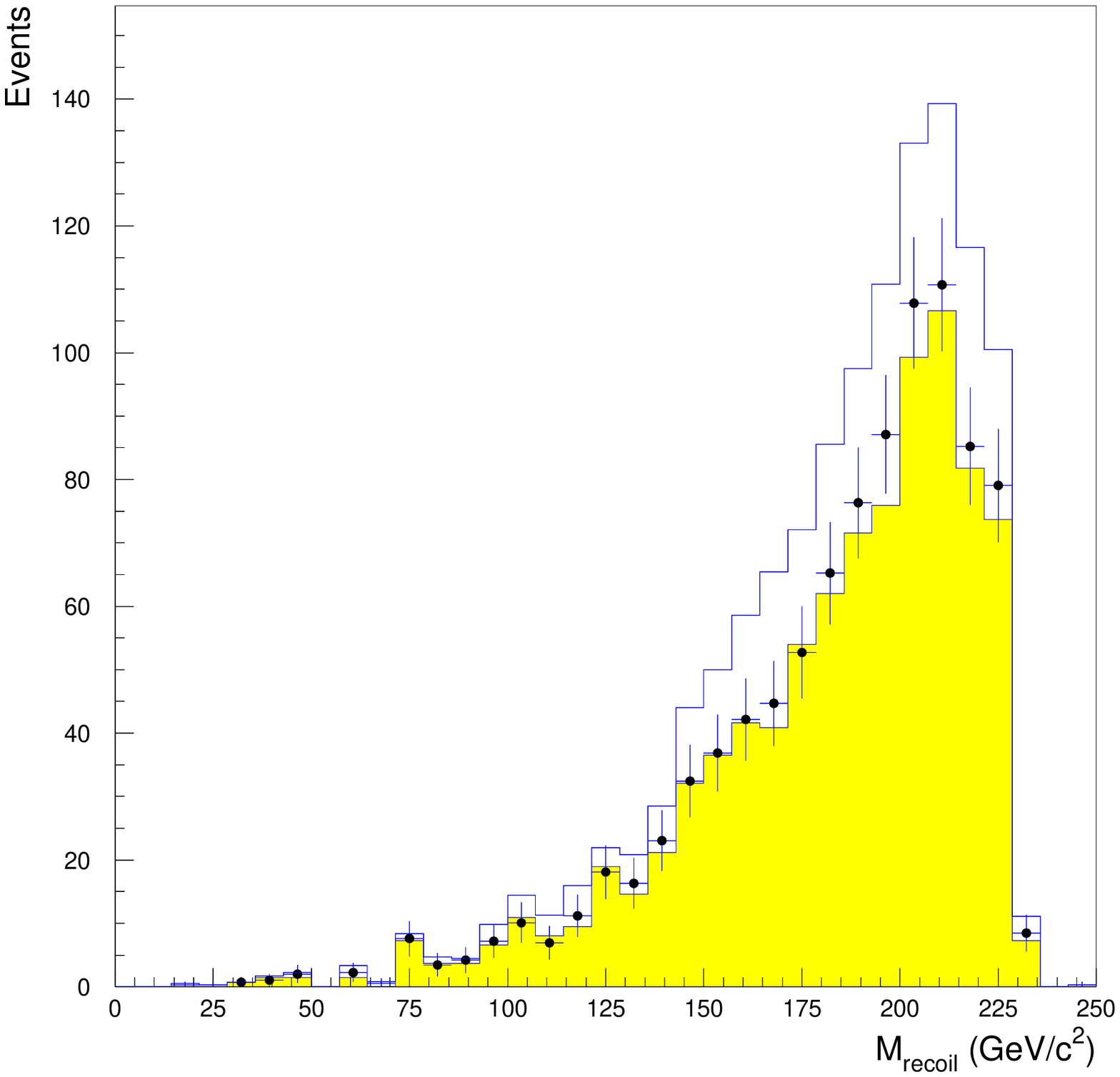,width=7.0cm,height=6.5cm} & 
\epsfig{file=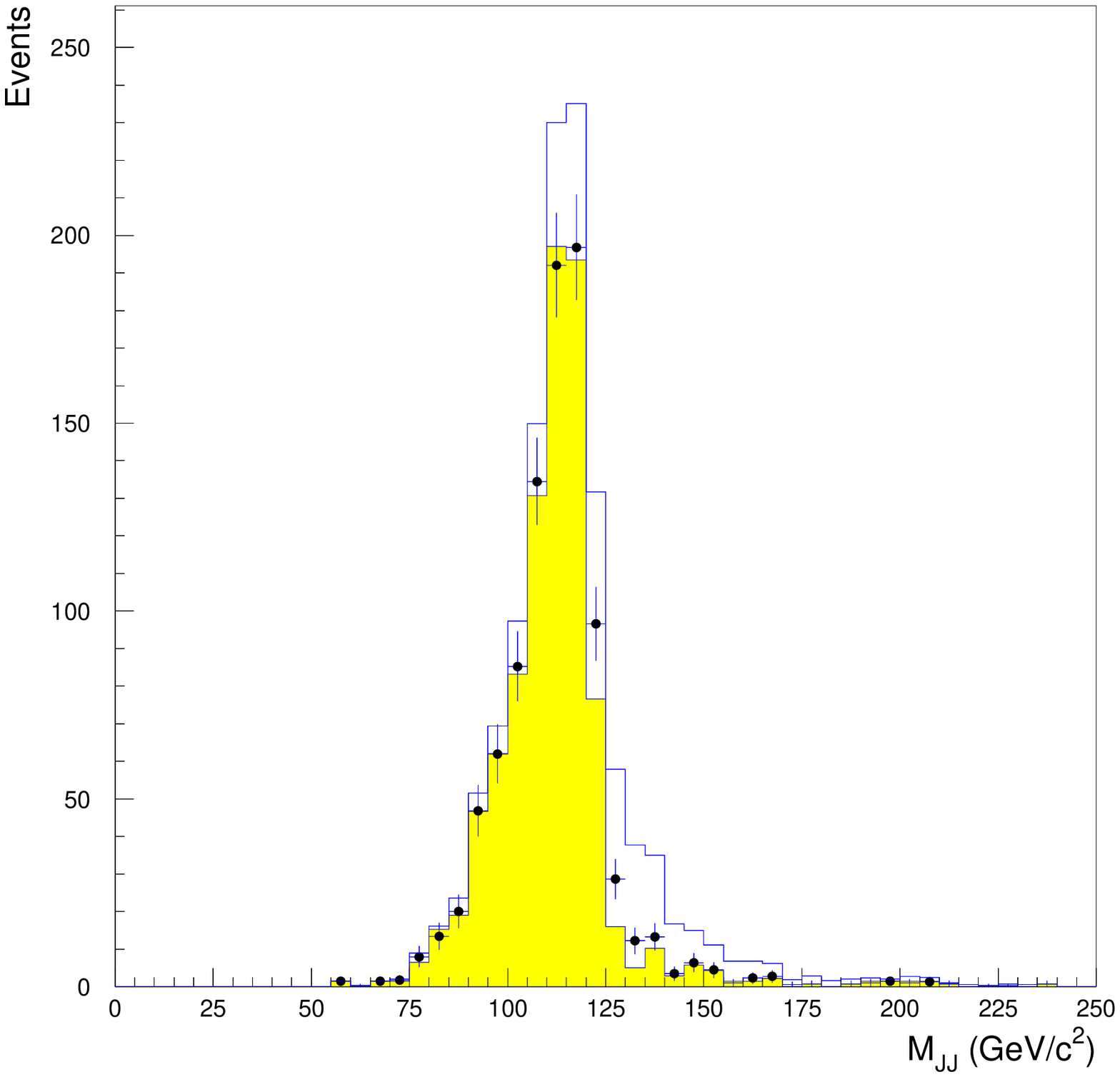,width=7.0cm,height=6.5cm} \\
\end{tabular}
\end{center}
\caption[]{\sl The distribution of the mass recoiling against the di-jet 
system (left) and the di-jet invariant mass (right) in $H \nu \bar \nu$ 
events. The open (shaded) histogram represents the reconstructed 
distributions with (without) $\gamma \gamma$ background and the points with
error bars represent the result after anti-$\gamma \gamma$ cut corrected by
the estimated efficiency loss.}
\label{fig:mass}
\end{figure}

\newpage
A $b$-tagging algorithm 
based on the impact parameter of the charged particles in each 
jets~\cite{btag} has been applied and the pair of $b$-tagged jets giving the 
mass closest to the 
nominal Higgs mass of 120~GeV/$c^2$ have been selected. 
Figure~\ref{fig:mass}
shows the reconstructed recoil mass to the two selected jets and their
invariant mass. A significant distortions towards higher mass values of the 
invariant mass distribution has been observed in presence of $\gamma \gamma$
background. However after applying the anti-$\gamma \gamma$ cut on 
$L_{\gamma \gamma}$ and rescaling the resulting distribution by the estimated
20\% reconstruction efficiency loss, the distributions agree well with those
expected for pure $H \nu \nu$ decays.

\section{Conclusions}

The effect of the overlap of $\gamma \gamma \rightarrow {\mathrm{hadrons}}$
to $H \nu \bar \nu$ events has been studied for the case of the {\sc Tesla}
$e^+e^-$ linear collider at $\sqrt{s}$ = 350~GeV, assuming Vertex Tracker
sensors able to resolve individual bunch crossing. 
The $H \nu \bar \nu$ channel, characterised by missing energy and 
$b$-tagged jets is expected to be particularly sensitive to the effects of 
$\gamma \gamma$ production. 
The $\gamma \gamma$ background to physics events can be substantially 
reduced, with moderate loss in reconstruction efficiency, by a combination of 
kinematical and vertex topology observables. The remaining background, being
confined to very forward hadron production, does not significantly interfere
with the mass reconstruction and cross-section measurement for Higgs studies
in the $WW$ fusion channel.

\end{document}